\documentclass[runningheads]{llncs}

\usepackage[T1]{fontenc}
\def\doi#1{\href{https://doi.org/\detokenize{#1}}{\url{https://doi.org/\detokenize{#1}}}}
\usepackage{graphicx}
\usepackage{listings}
\usepackage{amsmath}
\usepackage{amssymb}
\usepackage{graphicx}
\usepackage{bm}
\usepackage{color}
\usepackage{authblk}
\usepackage{booktabs}
\usepackage{float}
\usepackage{multirow}
\usepackage[colorlinks, citecolor=green]{hyperref}
\newcommand*\samethanks[1][\value{footnote}]{\footnotemark[#1]}
\begin{document}
\title{Learning-based and unrolled motion-compensated reconstruction for cardiac MR CINE imaging}
\titlerunning{Pre-print: Submitted and accepted by MICCAI 2022}
\authorrunning{Pre-print: Submitted and accepted by MICCAI 2022}

\author{}
\institute{}

\author{Jiazhen Pan\inst{1} \
Daniel Rueckert\inst{1,2} \
Thomas K\"ustner\inst{3}\thanks{contributed equally} \
Kerstin Hammernik\inst{1,2}\samethanks[1]
}
%index{Pan, Jiazhen}
%index{Rueckert, Daniel}
%index{K\"ustner, Thomas}
%index{Hammernik, Kerstin}

\institute{
Klinikum Rechts der Isar, Technical University of Munich, Germany
\and
Department of Computing, Imperial College London, United Kingdom
\and
Medical Image And Data Analysis (MIDAS.lab), University Hospital of Tübingen, Germany
}
\maketitle              

\begin{abstract}
% old one:
% Non-rigid motion of the heart in cardiac magnetic resonance (CMR) imaging remains a challenging problem for image reconstruction. Motion-compensated MR reconstruction (MCMR) is a powerful concept with considerable potential to solve this challenge. It consists of two coupled sub-problems: motion estimation (for known image) and image reconstruction (for known motion). A reliable reconstruction can only be achieved for accurately estimated motion. In this work, we propose a learning-based self-supervised MCMR framework and apply it in cardiac CINE. Contrary to conventional MCMR methods in which the motion is estimated at the beginning and remains unchanged during the whole iterative optimization process, we introduce a dynamic motion estimation process and embed it into the unrolled optimization. A temporally informed cardiac motion estimation network is applied and a joint optimization between the motion estimation and reconstruction is carried out. Experiments on 40 in-house acquired 2D CMR CINE datasets demonstrate that the proposed unrolled MCMR framework outperforms baseline methods (e.g. sequential motion estimation and reconstruction) substantially and its joint optimization mechanism is mutually beneficial for both sub-tasks (motion estimation and image reconstruction) especially when the MR image is highly undersampled.

Motion-compensated MR reconstruction (MCMR) is a powerful concept with considerable potential, consisting of two coupled sub-problems: Motion estimation, assuming a known image, and image reconstruction, assuming  known motion. In this work, we propose a learning-based self-supervised framework for MCMR, to efficiently deal with non-rigid motion corruption in cardiac MR imaging. Contrary to conventional MCMR methods in which the motion is estimated prior to reconstruction and remains unchanged during the iterative optimization process, we introduce a dynamic motion estimation process and embed it into the unrolled optimization. We establish a cardiac motion estimation network that leverages temporal information via a group-wise registration approach, and carry out a joint optimization between the motion estimation and reconstruction. Experiments on 40 acquired 2D cardiac MR CINE datasets demonstrate that the proposed unrolled MCMR framework can reconstruct high quality MR images at high acceleration rates where other state-of-the-art methods fail. We also show that the joint optimization mechanism is mutually beneficial for both sub-tasks, i.e., motion estimation and image reconstruction, especially when the MR image is highly undersampled.

% \keywords{Cardiac Magnetic Resonance Imaging \and Non-rigid registration network \and Motion-compensated image reconstruction.}
\end{abstract}
\section{Introduction}
% In the past decade, the increasing usage and the encouraging performance of cardiac magnetic resonance imaging (CMR) have demonstrated itself as a versatile technique competent for multiple tasks such as cardiac morphology and function assessment.
Cardiac magnetic resonance imaging (CMR) plays an essential role in evidence-based diagnostic of cardiovascular disease~\cite{von2017representation} and serves as the gold-standard for assessment of cardiac morphology and function~\cite{Lee2018}. High-quality cardiac image reconstruction with high spatial and temporal resolution is an inevitable prerequisite for this assessment. Shorter scan times with higher spatio-temporal resolution are desirable in this scenario. However, this requires high acceleration rates which in turn is only achievable if sufficient spatio-temporal information linked by the cardiac motion is shared during reconstruction. 

A large variety of CMR reconstruction methods have been dedicated to cope with cardiac motion during the reconstruction. They can be categorized into two sections: Implicit and explicit motion correction. Implicit motion correction during the reconstruction sidesteps the non-rigid cardiac motion estimation, which still remains one of the most challenging problems in CMR. Most research in this section focused either on exploiting spatio-temporal redundancies in complementary domains~\cite{k-t-focuss,CTFNet}, enforcing sparseness/low-rankness along these dimensions~\cite{Huang21,MALLRT,STORM}, or improving the spatio-temporal image regularization in an unrolled optimization model \cite{Kuestner2020CINENet,dl-espirit}. Yet in this sense, motion is only implicitly corrected without knowing or estimating the true underlying motion. 
% [JohannesSchmidt]
% On the other hand, explicit motion correction, incorporates motion information, for example obtained from image registration.

Motion can also be explicitly corrected during the reconstruction by applying motion estimation/registration models. The work of Batchelor et al.~\cite{Batchelor2005} has pioneered the field of motion-compensated MR reconstruction (MCMR). It proposed the idea of embedding the motion information as an explicit general matrix model into the MR reconstruction process and demonstrated some successful applications in leg and brain reconstruction. However, in CMR, the respiratory and cardiac motion is much more complex and therefore a more sophisticated motion model is required. Some endeavours were made first in the field of coronary magnetic resonance angiography (CMRA) to compensate respiratory motion~\cite{Prost2,Prost1} in which non-rigid motion models based on cubic B-Splines were employed~\cite{klein2009elastix,Modat2010}. However, these conventional registrations require substantial computation times in the order of hours making the practical implementation of image reconstruction infeasible. More recently, learning-based registration methods~\cite{voxelmorph1,voxelmorph2,Qi2021} have been proposed, which leverage trained neural networks to accelerate the estimation in inference. In~\cite{Qi2021-2} these learned registrations are embedded into a CMRA reconstruction for an unrolled CG-SENSE optimization~\cite{Pruessmann2001} with a denoiser regularizer~\cite{MoDL}. However, MCMR has been rarely studied in the context of cardiac CINE imaging~\cite{Cruz2021}. 

% Unlike the methods above in which the motion is pre-computed and assumed to be a fixed matrix during the whole optimization process while only the MR image is optimized, 
MCMR can be recast as two codependent and intertwined sub-optimization problems: Image reconstruction and motion estimation. This stands in contrast to methods in which only the MR image is optimized~\cite{Prost1,Prost2,Qi2021-2,Cruz2021}, whereas the motion is pre-computed and assumed to be fixed during the whole optimization process. A reliable reconstruction relies on precise motion estimation, while the accuracy of the motion prediction is subject to the quality of images. Odille et al.~\cite{Odille2008,Odille2016} introduced an iterative, alternating approach to solve this problem. Concurrently, \cite{CS_M,CS_M2} proposed to solve this joint optimization problem by leveraging variational methods. However, to the best of our knowledge, this problem has not been formulated and explored using deep learning-based approaches. Furthermore, as algorithm unrolling has been successfully applied in modern reconstruction approaches~\cite{MoDL,VN-Net} by learning image regularization from data, the possibility to unroll the MCMR optimization with motion estimation networks has not been explored.

%Furthermore, as unrolling the MR optimization with image denoiser networks is nowadays a common strategy for modern reconstruction methods\cite{VN-Net,MoDL}, the possibility to unroll the MCMR optimization with motion estimation networks has not been explored. 

In this work, we introduce a deep learning-based and unrolled MCMR framework to reconstruct CMR images in the presence of motion. The highlights of the work can be summarized as follows: First, we propose an unrolled MCMR optimization process with an embedded group-wise motion estimation network~\cite{Hammernik2021}. A joint optimization between the image reconstruction and motion estimation is proposed, while the motion estimation is updated iteratively with the reconstruction progress. %This mitigates the training difficulty, enables the network to estimate accurate motion in more challenging data with a higher undersampling rate and therefore accomplish a more meaningful reconstruction. 
Second, in contrast to~\cite{Prost2,Cruz2021,Qi2021-2} our proposed approach does not require a pre-computed motion field, which is usually obtained from initial non-motion compensated reconstruction. Finally, instead of reconstructing only one frame by compensating motion from all frames to one target frame, our proposed method provides a motion-resolved image sequence in which all frames are motion-corrected. 
The proposed framework was trained and tested on in-house acquired 2D CINE data of 40 subjects and compared to state-of-the-art MCMR methods. The conducted experiments demonstrate that the proposed method outperforms the competing methods especially in high acceleration situations, concluding that embedding the dynamic motion into an unrolled MCMR can drastically improve the reconstruction performance. 
% mitigate the training difficult -- network can be trained with higher R

\section{Methods}

Assume that $x^{(t)} \in \mathbb{C}^M$ is the $t$-th frame of the image sequence $x = [x^{(1)},\ldots,x^{(N)}]$ with $M$ pixels and $y^{(t)} \in \mathbb{C}^{MQ}$ denotes its corresponding $t$-th undersampled k-space data with $Q$ coils. In the context of CMR, the MCMR of the whole MR sequence with $N$ temporal frames can be formulated as
% \begin{align}
% \label{eq:normal_recon}
% \min\limits_{x} \sum_{t_1=1}^{N}\left\Vert \sum_{t_2=1}^{N} \mathbf{A}^{(t_2)} x^{(t_1)}  - y^{(t_1)} \right\Vert^2_2 %+\lambda\mathcal{R}(\mathbf{x}),
% \end{align}

\begin{align}
\label{eq:batchelor_recon}
\min\limits_{x, \mathbf{U}} \sum_{t_1=1}^{N}  \sum_{t_2=1}^{N} \left\Vert \mathbf{A}^{(t_2)} \mathbf{U}^{(t_1\rightarrow t_2)} x^{(t_1)}  - y^{(t_2)} \right\Vert^2_2, %+ \lambda\mathcal{R}(\mathbf{x}) ,
\end{align}
% \begin{align}
% \label{eq:batchelor_recon}
% \min\limits_{x} \sum_{t=1}^{N} \left\Vert \mathbf{A} \mathbf{U}^{(t)} x^{(t)}  - y\right\Vert^2_2, %+ \lambda\mathcal{R}(\mathbf{x}) ,
% \end{align}
%
% \begin{align}
% \label{eq:normal_recon}
% \min\limits_{\mathbf{x}} \left\Vert \sum_{t=1}^{N} \mathbf{A}^{(t)} \mathbf{x}^{all}  - y^{(t)} \right\Vert^2_2,
% \end{align}
where $\mathbf{A}^{(t)}$ denotes the MR multi-coil encoding operator with $\mathbf{A}^{(t)}=\mathbf{D}^{(t)}\mathbf{F}\mathbf{C}$ for coil-sensitivity maps $\mathbf{C}$, Fourier transform $\mathbf{F}$ and the undersampling matrix $\mathbf{D}$. Based on the idea of \cite{Batchelor2005},  we build a motion matrix $\mathbf{U}^{(t_1\rightarrow t_2)}$ representing the warping from the temporal frame $t_1$ to $t_2$, which is obtained from the estimated motion fields $u^{(t_1\rightarrow t_2)}$ (introduced in Section~\ref{network}). 
% from the estimated motion fields $u^{(t_1\rightarrow t_2)}$, which represents the warping from the temporal frame $t_1$ to $t_2$. 

% into Eq. \ref{eq:normal_recon}:
%\begin{align}
%\label{eq:batchelor_recon}
%\min\limits_{x, \mathbf{U}} \sum_{t_1=1}^{N}\left\Vert \sum_{t_2=1}^{N} \mathbf{A}^{(t_2)} \mathbf{U}^{(t_1\rightarrow t_2)} x^{(t_1)}  - y^{(t_1)} %\right\Vert^2_2 + \lambda\mathcal{R}(\mathbf{x}) ,
%\end{align}
% \begin{align}
% \label{eq:batchelor}
% x_{i+1} = \arg \min\limits_{\mathbf{x}_i} \left\Vert \sum_{t=1}^{N} \mathbf{A}^{(t)} \mathbf{U}^{(all\rightarrow t)} \mathbf{x}^{all}_i  - y^{(t)} \right\Vert^2_2.
% \end{align}
In traditional MCMR~\cite{Prost2,Cruz2021,Qi2021-2}, only the images $x$ are optimized, whereas the motion is pre-computed from initial (non-motion compensated) reconstructed images just for once and assumed to be constant during the whole optimization process. However, motion estimation on the initial motion-corrupted and artifact-degraded images can be inaccurate, incurring error-propagation during the optimization. % Motivated by the success of using a joint model~\cite{Odille2016}, the joint optimization for image $\mathbf{x}$ and motion $\mathbf{U}$ as described in Eq.~\ref{eq:batchelor_recon} should be superior.
Therefore, a joint optimization as proposed in~\cite{Odille2016} for image $x$ and motion $\mathbf{U}$ as described in Eq.~\ref{eq:batchelor_recon} is desired. 

\subsection{Motion-compensated image reconstruction framework}
\begin{figure}[!t]
    \centering
    \includegraphics[width=\linewidth]{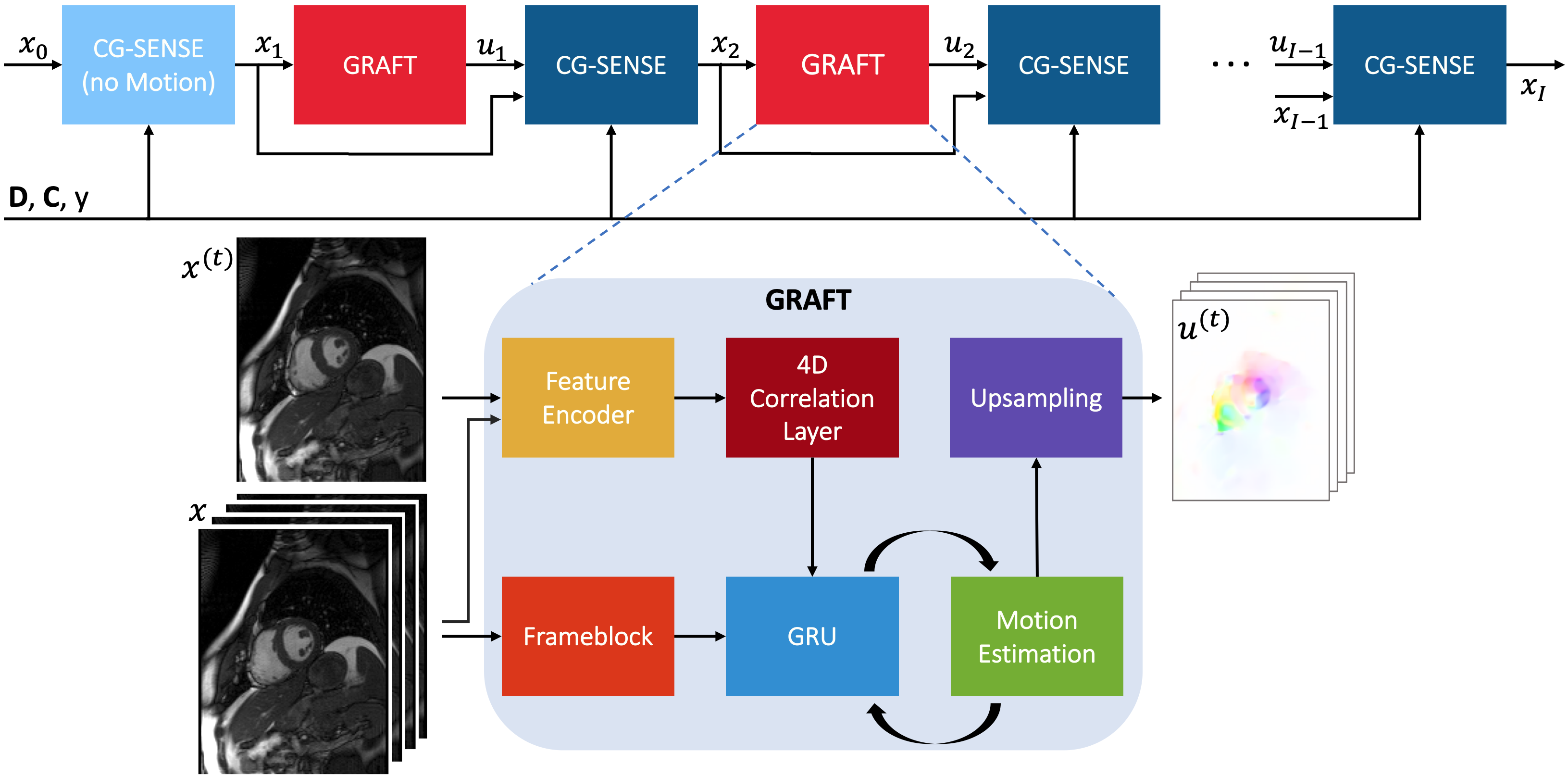}
    \caption{The proposed unrolled MCMR framework. A dynamic joint optimization is performed between the image reconstruction in the CG-SENSE block and the motion correction from the motion estimator GRAFT. GRAFT is a group-wise motion estimation network leveraging the temporal redundancy to conduct a 1-to-$N$ 
    % $u^{(t)} = \left[ u^{(t\rightarrow 1)}, \ldots,  u^{(t\rightarrow N)} \right]$
    motion estimation. This estimation is carried out $N$ times to accomplish an $N$-to-$N$ motion correction for all temporal frames.}
    \label{fig:architecture}
\end{figure}
% Motivated by the joint model~\cite{Odille2016} to solve Eq.~\ref{eq:batchelor_recon}, 
In this work we unroll the MCMR in Eq.~\ref{eq:batchelor_recon} with a motion estimation network and apply the joint optimization between the image reconstruction and motion estimation. The framework is illustrated in Fig.~\ref{fig:architecture}. First, the zero-filled undersampled image sequence $x_{0}$ is fed to the conjugate-gradient SENSE block~\cite{Pruessmann2001} along with $y$, $\mathbf{C}$, and $\mathbf{D}$. The first SENSE block conducts a reconstruction without motion embedding. Afterwards, the reconstructed image sequence $x_1$ is passed to a motion estimation network (GRAFT, introduced in Section \ref{network}) as inputs and the first motion sequence $u_{1}$ comprising of all pairs of motion frames is predicted. This motion sequence is then applied to the input together with $y$, $\mathbf{C}$ and $\mathbf{D}$ in the next CG-SENSE block, while the previous reconstruction is used as a regularization. These CG-SENSE blocks solve Eq.~\ref{eq:batchelor_recon} with an additional $\ell_2$ regularization by the image from the previous step and output a motion-corrected image sequence with temporarily freezing $\mathbf{U}_{i-1}$ and $x_{i-1}$, following %Eq.~\ref{eq:reconimage}. 
\begin{align}
    x_i = \arg\min\limits_{x} \sum_{t_1=1}^{N}  \sum_{t_2=1}^{N} \left\Vert \mathbf{A}^{(t_2)} \mathbf{U}_{i-1}^{(t_1\rightarrow t_2)} x^{(t_1)}  - y^{(t_2)} \right\Vert^2_2 + \frac{1}{2\lambda}\left\Vert x - x_{i-1}\right\Vert^2_2.
    \label{eq:reconimage}
\end{align}
%
% The estimation difficulty at the first iteration is the highest...
This alternating scheme is carried out for $I$ iterations. The motion estimation difficulty of GRAFT is alleviated gradually with the progress of this alternating scheme along with the image-quality improvement.

\subsection{Motion Estimation Network}\label{network}
There exists various motion estimation approaches in the field of medical imaging. In this work, we select GRAFT~\cite{Hammernik2021} for motion estimation, which takes the full image sequence $x$ together with the target frame $x^{(t)}$ as inputs to predict the motion sequence $u^{(t)} = \left[ u^{(t\rightarrow 1)}, \ldots,  u^{(t\rightarrow N)} \right]$ for all $N$ frames. GRAFT is suitable to be embedded into our framework due to its accuracy, speed and efficiency.

\textit{Accuracy} An accurate motion estimation is the essential requirement for the unrolled MCMR framework. GRAFT contains a 4D correlation layer and iterative Gated Recurrent Unit (GRU) blocks to conduct precise motion predictions. It has shown superior results compared to conventional registration methods~\cite{Hammernik2021}. Furthermore, it is a group-wise estimation network considering all temporal frames and has a dedicated frameblock leveraging temporal redundancies to mitigate the impact of through-plane-motion. Because of this group-wise attribute, temporal coherence can be instilled during the training by appending temporal regularization in the loss function.
%A temporal regularization in the loss function reinforces its group-wise attribute.
%

\textit{Speed} Since the motion registration is invoked at each iteration step and for all pairs of motion frames, a fast motion estimation method is required. The reconstruction of $N$ frames involves $N^2$ motion predictions. In this context, traditional registration-based methods such as~\cite{klein2009elastix,vercauteren2009diffeomorphic} require hours to perform $N^2$ calculations, rendering them impractical for our framework. In contrast, GRAFT only requires a few seconds to estimate the motion of all pairs.%, making the practical implementation accessible.

\textit{Efficiency}
% As mentioned above, the reconstruction of $N$ frames entails $N^2$ calculations, and over multiple iterations. 
For end-to-end learning over all iterative stages, the losses and gradients need to be accumulated. In the context that $N^2$ calculations (and for multiple iterations) are entailed, the training of large network architectures with more than 20 million trainable parameters~\cite{voxelmorph2,Qi2021} becomes infeasible due to their vast GPU memory footprint (>48GB). In contrast, GRAFT has only 5 million trainable parameters circumventing the GPU overcharge problems.

In order to use GRAFT for motion estimation and integrate it in our unrolled framework we define the following loss function to train GRAFT:
\begin{equation}
\begin{aligned}
\mathcal{L}_i= \sum_{t_1=1}^{N}\sum_{t_2=1}^{N}  &\left\Vert\rho  \left( \mathbf{U}_i^{(t_1\rightarrow t_2)} x_{gt}^{(t_1)}-x_{gt}^{(t_2)}\right)\right\Vert_1 \\
+&\alpha \sum_{t_1=1}^{N}\sum_{t_2=1}^{N} \sum_{d \in x, y} \left\Vert\nabla_{d} u_i^{(t_1 \rightarrow t_2)}\right\Vert_1 +\beta \sum_{t_1=1}^{N} \left\Vert\nabla_{t} u_i^{(t_1)}\right\Vert_1.
\end{aligned}
\end{equation}
The first term ensures data fidelity during the training. We warp the ground-truth frame $x_{gt}^{(t_1)}$ to the target ground-truth frame $x_{gt}^{(t_2)}$ using the motion $u_i^{(t_1\rightarrow t_2)}$, which is predicted from GRAFT based on the previously reconstructed image $x_{i-1}$. It should be noted that we apply the ground-truth frame in the loss function to train GRAFT, so that GRAFT is forced to learn and extract genuine correlations from the undersampled images and that any motion falsely originating from image artifacts are not rewarded. The Charbonnier function $\rho(x)=(x^2+10^{-12})^{0.45}$~\cite{Sun2018,Pan2021} is employed as the penalty function. Additional regularization terms for the spatial plane weighted by $\alpha$ and along the temporal axis weighted by $\beta$ are included in the loss to mitigate estimation singularities and to ensure motion coherence in spatial and temporal domain. This loss is calculated after every unrolled motion estimation step with progressive exponential weight decay $\mathcal{L}_{total}=\sum_{i=1}^{I} \gamma^{I-i}\mathcal{L}_{i}$. %where $\gamma=0.6$ during training.

\section{Experiments}
Training was carried out on 35 subjects (a mix of patients and healthy subjects) of in-house acquired short-axis 2D CINE CMR, whereas testing was performed on 5 subjects. Data were acquired with 30/32/34 multiple receiver coils and 2D balanced steady-state free precession sequence on a 1.5T MR (Siemens Aera with TE=1.06 ms, TR=2.12 ms, resolution=$1.9{\times}1.9\text{mm}^2$ with 8mm slice thickness, 8 breath-holds of 15s duration). Image sequence size varies from $176{\times}132$ (smallest) to $192{\times}180$ (largest) with 25 temporal cardiac phases. A stack of 12 slices along the long axis was collected, resulting in 304/15 image sequence (2D+t) for training/test, while the apical slices without clear cardiac contour were excluded. 

The proposed framework with GRAFT as the embedded motion estimator was trained on an NVIDIA A40 GPU with AdamW~\cite{adamw} (batch size of 1, learning rate 1e-4 and weight decay of 1e-5). The number of unrolled iterations is set to 3 during training, but this can be flexibly adapted during inference and iterations are stopped when the peak signal to noise ratio (PSNR) converges (PSNR increment $<0.1$). The trainable weights of GRAFT are shared during the iterative optimization. The hyperparameters $\alpha$, $\beta$, $\gamma$ and $\lambda$ were set to 10, 10, 0.6 and 2 respectively. Three trainings and tests were conducted separately on retrospectively undersampled images with VISTA~\cite{Ahmad2015} of R = 8, 12 and 16 acceleration without any prior reconstruction or prior motion correction as conducted in~\cite{Prost2,Cruz2021,Qi2021-2}. During the training and testing, raw multi-coil k-space data were used for the reconstruction.
% Furthermore, we used Tikhonov regularization~\cite{Tikhonov} during reconstruction to regularize the image sequence in x-y plane.
% We compare the proposed approach with non-motion compensated CG-SENSE (N-CG-SENSE) reconstruction and a motion-compensated CG-SENSE (i.e. non-iterative MCMR method) in which the motion is pre-computed and then fed to the CG-SENSE reconstruction. For the pre-computed motion estimation in MC-CG-SENSE, GRAFT (alone without this unrolling framework) and Elastix~\cite{klein2009elastix} are chosen as the comparative baseline. 
We compare the proposed approach with non-motion compensated CG-SENSE (N-CG-SENSE) reconstruction and with non-iterative MCMR methods. In the latter case, the motion is pre-computed from GRAFT~\cite{Hammernik2021} and Elastix~\cite{klein2009elastix} and set to be constant during the whole optimization process.

\section{Results and Discussion}
\begin{figure}[H]%[!t]
    \centering
    \includegraphics[width=\linewidth]{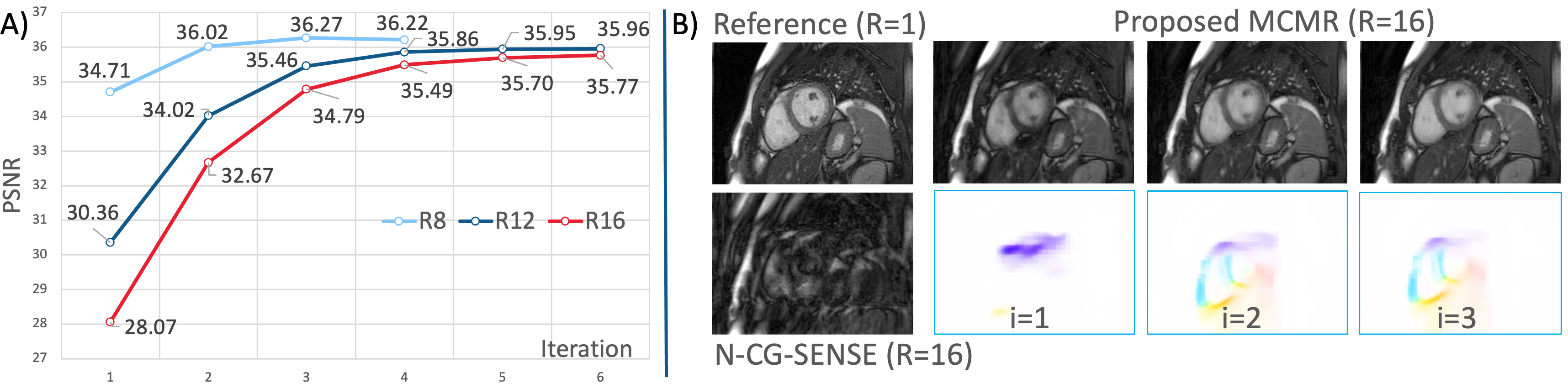}
    \caption{Quantitative and qualitative results of proposed MCMR method. \textbf{A}) Performance measured by peak signal to noise ratio (PSNR) for increasing unrolled iteration numbers over all test subjects and temporal frames for retrospective undersampling with R=8, 12 and 16. \textbf{B}) Qualitative image quality of the proposed MCMR at R=16 over the first three iterations (i=1,2,3) in comparison to the non-motion compensated CG-SENSE (N-CG-SENSE) at R=16 and the fully-sampled reference image (R=1).}
    \label{fig:Iteration}
\end{figure}

Fig.~\ref{fig:Iteration}A reveals the relation between the estimation performance (indicated by averaged PSNR over all test subjects and temporal frames) and the unrolled iteration number of the proposed framework during inference for R=8, 12 and 16. %\footnote{We observed that during the evaluation on the R=16, experiment of training on R=12 CINE data has overall better results than the training on R=16. Therefore, we use the experiment of training on R=12 and evaluation on R=16 CINE data to present the R=16 evaluation hereafter.}. 
The reconstruction accuracy is improved with increasing iteration numbers for all three cases. The reconstruction for R=12 and R=16 benefit most from our framework and obtain a significant performance lifting for the first three iterations. The advantage of unrolling the motion estimator in the joint MCMR optimization process is qualitatively illustrated in Fig.~\ref{fig:Iteration}B, in which the MR image is highly undersampled with R=16. Not only is the image quality improved with the course of the iteration, but the motion estimator can also deliver more precise and detailed estimation by virtue of the higher quality image. A full qualitative analysis with R=8 and 12 is shown in Supplementary Fig.~S1.
% This joint optimization is crucial for the reconstruction case of a highly undersampled image as shown in the case of R=16 in Fig.~\ref{fig:Iteration} right part.  
Furthermore, the motion estimation based on unprocessed and artifact-affected images (from high undersampling rates) is challenging. We also trained GRAFT without the proposed unrolled optimization framework~\cite{Hammernik2021} to generalize for motion estimation from images with different undersampling rates. However, training started to fail for accelerations R=12 and beyond. In contrast, training difficulty is reduced if we embed GRAFT into the proposed joint optimization framework while the image quality is restored with the progress of the optimization. This makes GRAFT capable of estimating meaningful motion as showcased in Supplementary Fig.~S1 for R=12 and even for R=16. Moreover, this mechanism also introduces a data augmentation process while more high-quality data is generated during training.

% \begin{figure}[!t]
%     \centering
%     \includegraphics[width=\linewidth]{PSNR_Iteration.png}
%     \caption{Influence of choosing the proposed framework's unrolled iteration number on the reconstruction accuracy during the evaluation for different acceleration rates R.}
%     \label{fig:PSNR_Iteration}
% \end{figure}
\begin{figure}[!t]
    \centering
    \includegraphics[width=\linewidth]{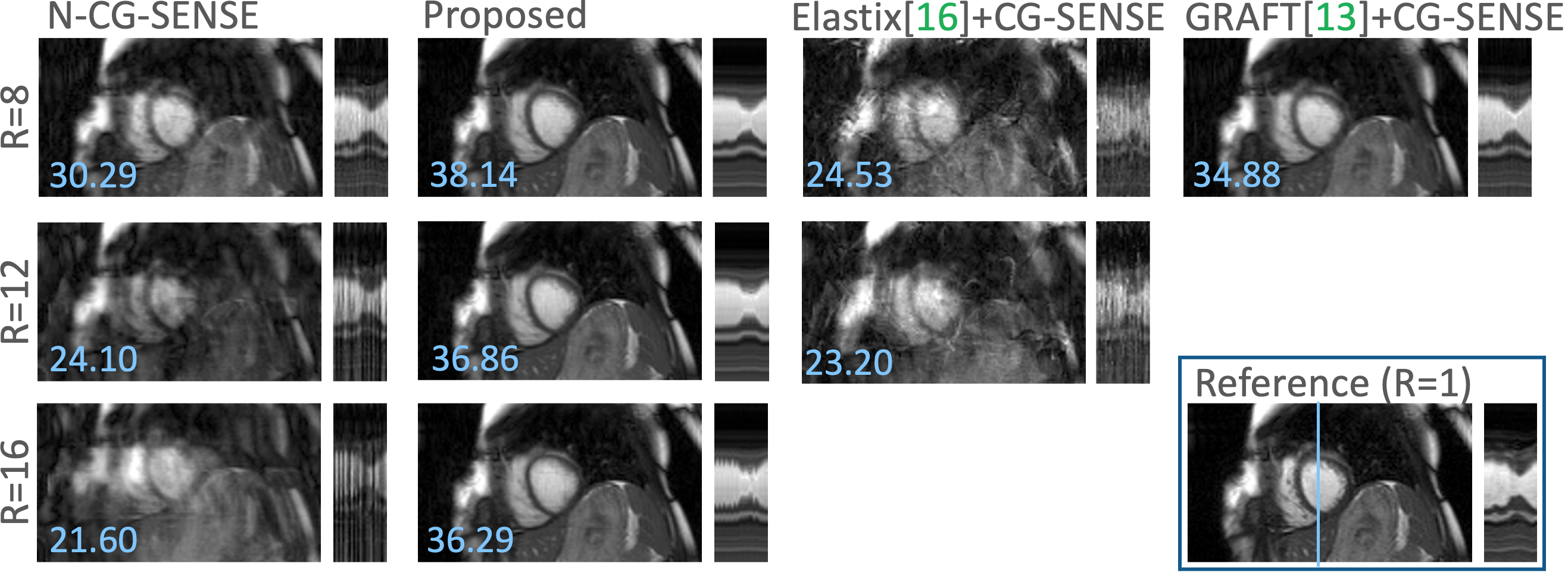}
    \caption{Qualitative reconstruction performance of CINE CMR in spatial (x-y) and spatio-temporal (y-t) plane of non-motion compensated CG-SENSE (N-CG-SENSE), proposed MCMR at third iteration, CG-SENSE with pre-estimated motion from GRAFT and Elastix, for different acceleration rates R=8, 12 and 16 in comparison to the fully-sampled (R=1) reference. The selected superior-inferior (y-axis) is marked with a blue line in the reference image. The obtained PSNR in comparison to the fully-sampled reference is shown at the bottom left. The highly accelerated cases for Elastix/GRAFT + CG-SENSE did not converge therefore results are not shown.}
    \label{fig:Comparison}
\end{figure}

A comparison study of the proposed MCMR to N-CG-SENSE and CG-SENSE reconstruction with pre-computed motion based on GRAFT and Elastix is shown in Fig.~\ref{fig:Comparison}. The training of GRAFT alone failed for R=12 and 16, whereas motion estimation with Elastix for R=16 was not conducted due to the poor quality results collected at R=12. The results are shown after the third iteration for the proposed MCMR, providing an optimal trade-off between time and performance. Our approach restores the undersampled MR sequence with high quality without artifacts in spatial and temporal domain. A quantitative analysis over all test subjects including reconstruction of all 25 temporal frames is shown in Table \ref{tab:QuantComp}. Although the execution time of our approach is longer than the original GRAFT if the iteration is set $>1$, an overall more precise reconstruction is achieved. Furthermore, the overall reconstruction is within acceptable clinical durations. The proposed MCMR enables to reconstruct an undersampled image with a much higher acceleration rate and offers the high flexibility to perform the reconstruction depending on different time/performance requirement.

%the advantages it brings including an overall more precise reconstruction, the possibility to reconstruct an undersampled image with a much higher acceleration rate, and the high flexibility to perform the reconstruction depending on different time/performance requirement are irresistible. 
\begin{table}[!t]
    \centering
    \caption{Quantitative analysis of reconstruction for accelerated CINE CMR (R=8, 12 and 16) using the proposed MCMR method, non-motion compensated CG-SENSE (N-CG-SENSE), GRAFT+CG-SENSE and Elastix+CG-SENSE. Peak signal-to-noise ratio (PSNR) and structural similarity index (SSIM)~\cite{Wang2004} are used to evaluate all test subjects. Their mean value, standard deviations are shown next to the respective methods execution times. The best results are marked in bold. The failed or inferior experiments are marked with 'N.A.'.}
    \begin{tabular}{ccccc}
        \toprule
        Acc R & Methods & SSIM & PSNR & Time (s)\\ \midrule
        \multirow{4}*{8} 
        & Proposed MCMR & \textbf{0.943} (0.018) & \textbf{36.26} (2.22) & 18.81s\\
        ~ & GRAFT~\cite{Hammernik2021} + CG-SENSE & 0.913 (0.019) & 34.93 (1.80) & 6.27s\\
        ~ & Elastix~\cite{klein2009elastix} + CG-SENSE & 0.645 (0.057) & 25.04 (2.10) & 4281s\\
        ~ & N-CG-SENSE & 0.821 (0.038) & 30.80 (2.15) & 1.37s\\  \midrule
        \multirow{4}*{12} 
          & Proposed MCMR & \textbf{0.932} (0.018) & \textbf{35.45} (2.00) & 18.81s \\
        ~ & GRAFT + CG-SENSE & N.A. & N.A. & 6.27s \\
        ~ & Elastix + CG-SENSE & 0.568 (0.072) &  23.51 (2.20) & 4281s \\
        ~ & N-CG-SENSE & 0.637 (0.062) & 24.40 (2.39) & 1.37s \\ \midrule
        \multirow{4}*{16} 
          & Proposed MCMR & \textbf{0.927} (0.019) & \textbf{34.78} (1.86)  & 18.81s \\
        ~ & GRAFT + CG-SENSE & N.A. & N.A. & 6.27s \\
        ~ & Elastix + CG-SENSE & N.A. & N.A. & 4281s \\
        ~ & N-CG-SENSE & 0.531 (0.08) & 21.736 (2.45) & 1.37s \\ \bottomrule
    \end{tabular}
    \label{tab:QuantComp}
\end{table}

The proposed MCMR framework also has some limitations: Currently, the proposed MCMR cannot guarantee that the estimated motion is diffeomorphic. The possibility to integrate the scaling and squaring layer~\cite{Diffeo} to ensure diffeomorphic motion estiamtion will be investigated in future work. Furthermore, we have not studied the interaction between our explicit motion correction framework with implicit motion correction (denoising) networks, which is also subject to future work. Moreover, we plan to evaluate the performance of our method on prospectively undersampled data in the future. Finally, the hyper-parameters applied in this work are estimated empirically and they might not be the optimal combination. Integration of the learnable hyper-parameters tuning~\cite{Hyper} will be considered in our next step.

% Furthermore, we have not embedded a denoising network as an image regularizer into our framework. Our future work will also look into the potential interaction between motion estimation and a learnable image denoiser

% Finally, during experiments we find that sometimes a reconstruction shows inferior result with artefacts, while still score an acceptable PSNR/SSIM. A more reliable metric to reflect the quality of MR reconstruction need to be found.

\section{Conclusion}
In this work, we proposed a deep learning-based MCMR framework and studied it in CINE CMR imaging. We explored the possibility to unroll the MCMR optimization with a motion estimation network while a dynamic joint optimization between the reconstruction and the motion estimation is carried out. Although this idea is still at a nascent stage, its potential in high quality reconstruction of highly accelerated data can be appreciated. The conducted experiments against baseline methods showcased a rapid, more robust and more precise reconstruction of our proposed framework.

\section{Acknowledgements}
This work was supported in part by the European Research Council (Grant Agreement no. 884622).

%its potential is showcased in the conducted experiments compared to the baseline methods with a rapid, more robust and more precise reconstruction.

%its potential is showcased, enabling high quality reconstruction of highly accelerated data. The proposed approach showed in comparison to the baseline methods a rapid, more undersample-robust and more precise reconstruction.

% \subsubsection{Acknowledgements} Please place your acknowledgments at
% the end of the paper, preceded by an unnumbered run-in heading (i.e.
% 3rd-level heading).

\bibliographystyle{splncs04}
\bibliography{bibliography.bib}

\end{document}

\title{Supplementary Materials}
\titlerunning{ }
\author{}
\institute{}
\maketitle              
%
%
%
% \section{Introduction}

% \renewcommand{\thefigure}{S\@arabic\c@figure}
% \renewcommand{\thetable}{S \arabic{figure}}
\renewcommand\thefigure{S\arabic{figure}}   
% \thefigure{\thesection.\arabic{Supplementary Figure}}
\begin{figure}[h]
    \centering
    \includegraphics[width=\linewidth]{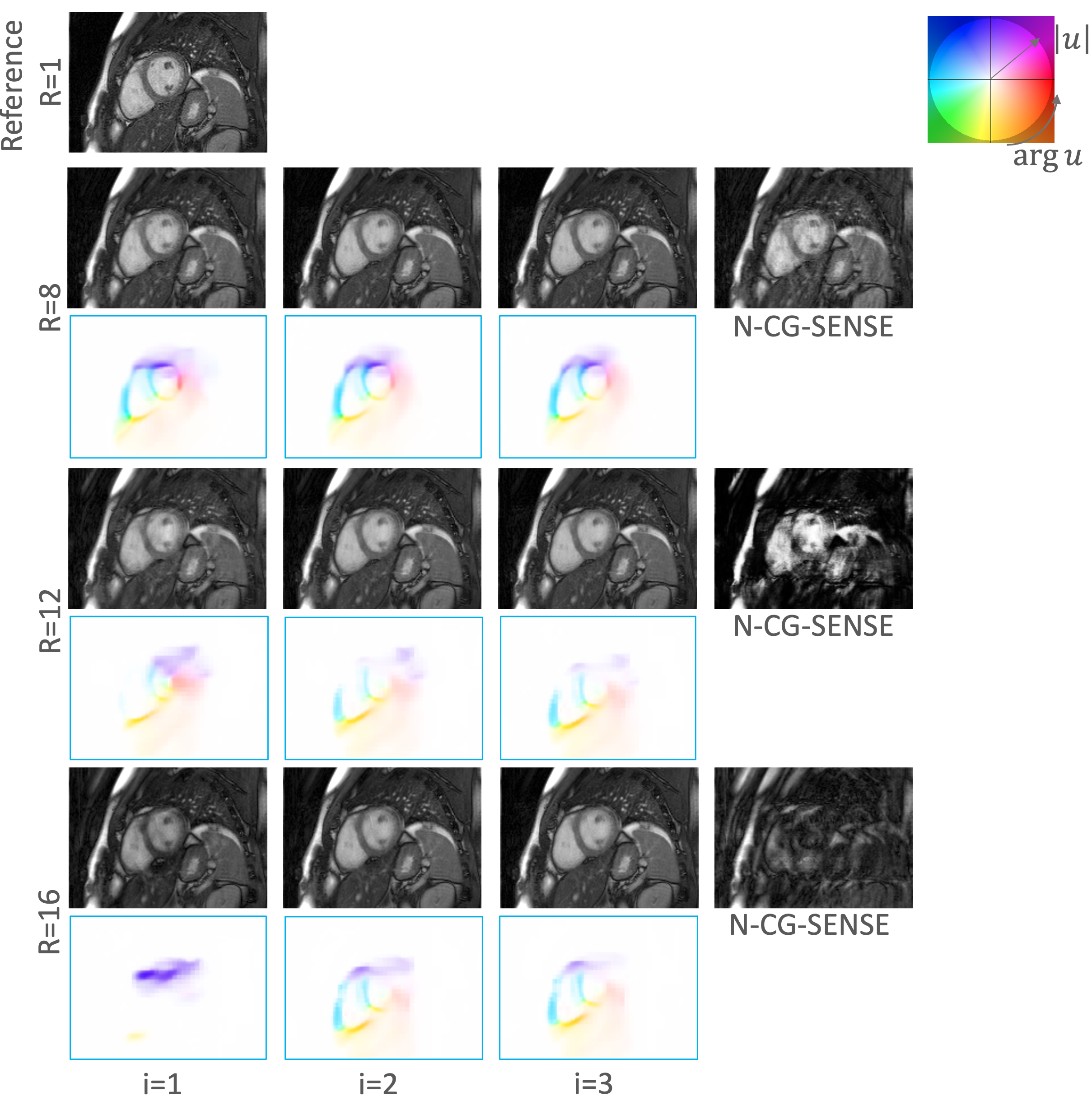}
    \caption{Reconstruction and motion estimation performance of the proposed MCMR method at R=8, 12 and 16 acceleration in the first three iterations (i=1,2,3) during inference. The non-motion compensated CG-SENSE (N-CG-SENSE) and the reference image (R=1) are shown as comparison.}
\end{figure}